\documentclass[
reprint,
superscriptaddress,
amsmath,amssymb,
aps, prl, floatfix,
]{revtex4-1}

\usepackage{graphicx}
\usepackage{dcolumn}
\usepackage{bm}

\newcommand{\abs}[1]{\left|{#1}\right|}
\usepackage{xcolor}

\begin{document}

\preprint{APS/123-QED}

\title{Spatial polarization gating of
high-harmonic generation in solids}

\author{Pieter J. van Essen}
\email{p.vessen@arcnl.nl}
\affiliation{Advanced Research Center for Nanolithography,\\ Science Park 106, 1098 XG,\\ Amsterdam, The Netherlands}%
\author{Brian de Keijzer}
\affiliation{Advanced Research Center for Nanolithography,\\ Science Park 106, 1098 XG,\\ Amsterdam, The Netherlands}%
\author{Tanya van Horen}
\affiliation{Advanced Research Center for Nanolithography,\\ Science Park 106, 1098 XG,\\ Amsterdam, The Netherlands}%
\author{Eduardo B. Molinero}
\affiliation{Instituto de Ciencia de Materiales de Madrid,\\ Consejo Superior de Investigaciones Científicas (ICMM-CSIC),\\ E-28049 Madrid, Spain}%
\author{Álvaro Jiménez Galán}
\affiliation{Instituto de Ciencia de Materiales de Madrid,\\ Consejo Superior de Investigaciones Científicas (ICMM-CSIC),\\ E-28049 Madrid, Spain}%
\affiliation{Max-Born-Institute for Nonlinear Optics and Short Pulse Spectroscopy, \\ Max-Born-Strasse 2A,\\ D-12489 Berlin, Germany}
\author{Rui. E.F. Silva}
\affiliation{Instituto de Ciencia de Materiales de Madrid,\\ Consejo Superior de Investigaciones Científicas (ICMM-CSIC),\\ E-28049 Madrid, Spain}%
\affiliation{Max-Born-Institute for Nonlinear Optics and Short Pulse Spectroscopy, \\ Max-Born-Strasse 2A,\\ D-12489 Berlin, Germany}
\author{Peter M. Kraus}%
\email{p.kraus@arcnl.nl}
\affiliation{Advanced Research Center for Nanolithography,\\ Science Park 106, 1098 XG,\\ Amsterdam, The Netherlands}%
\affiliation{Department of Physics and Astronomy, and LaserLaB, Vrije Universiteit,\\ De Boelelaan 1105, 1081 HV \\ Amsterdam, The Netherlands}

\date{\today}

\begin{abstract}
High-harmonic generation from solids can be utilized as probe of ultrafast dynamics, but thus far only over extended sample areas, since its spatial resolution is diffraction-limited.
Here we propose spatial polarization gating, that is using a spatially varying ellipticity of a driving laser pulse to reduce the spatial profile of high-harmonic emission below the diffraction limit and hence increase spatial resolution. We show experimentally and by numerical simulations that our method is generally applicable as suppressing high harmonics in elliptical fields is a common response in all solids.
We also briefly explore the possibility of applying this technique to widefield imaging, specifically to nonlinear structured illumination microscopy. Our findings indicate that spatial polarization gating can enable all-optical femto-to-attosecond label-free imaging beyond the Abbe limit.
\end{abstract}


\maketitle
\begin{figure*}
    \centering
    \includegraphics[width=\linewidth]{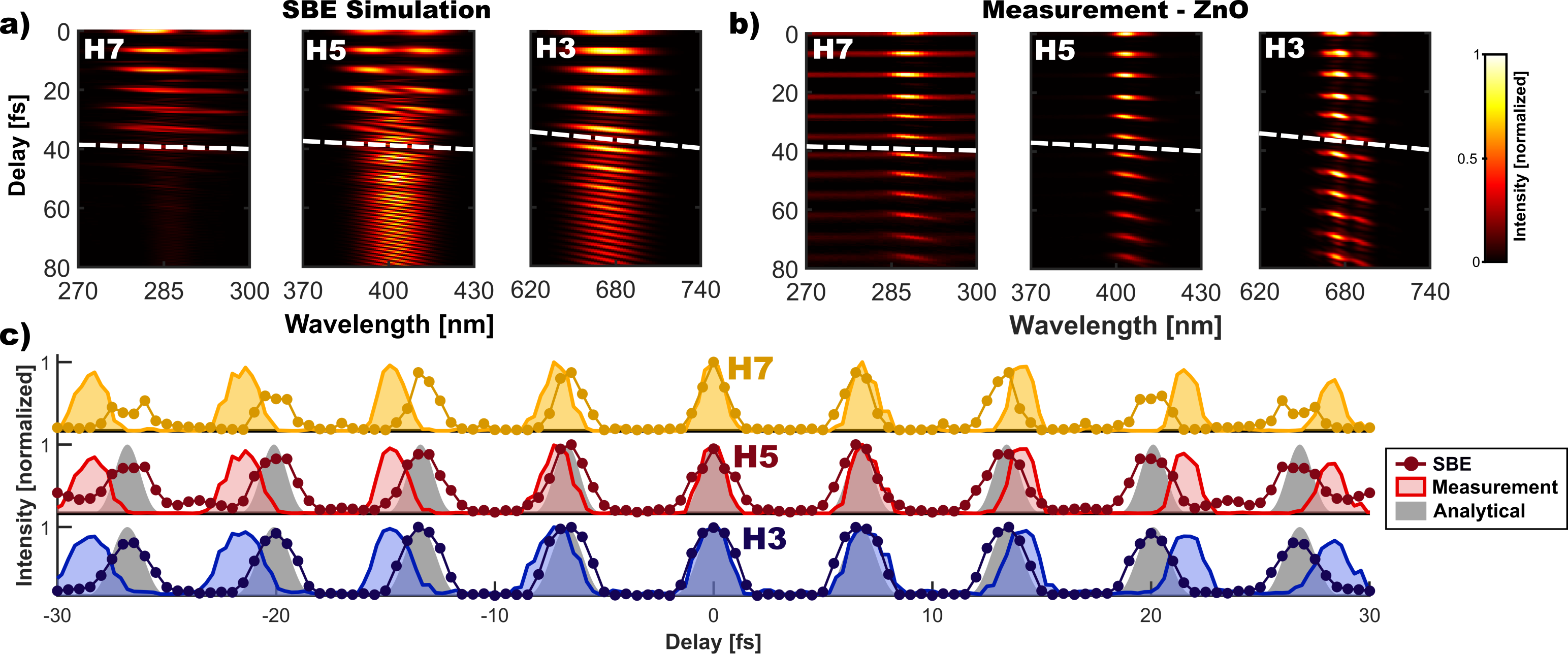}
    \caption{
    (a) SBE simulations and (b) experiments in ZnO for harmonics H3-H7 show an intensity modulation as function of changing delay (phase) between two 2000-nm orthogonally polarized (s- and p-polarized) pulses. This scan maps delay to ellipticity.  HHG emission in the diagonal polarization direction is selected and shown here. The white dashed lines all have the same slope and indicate the spectral tilt. (c) Integrated harmonic intensity for both the SBE simulation and measurement. For H3 and H5 also the analytical results using equation \ref{eq:analyticalModel} with the measured suppression parameters are shown.  
    }
    \label{fig:SBE_Delay}
\end{figure*}
A plethora of recent works have demonstrated high-harmonic generation (HHG) as a highly sensitive probe for ultra-fast phenomena in solids \cite{Heide2022coherence,Hohenleutner2015,Zhang2024,Bionta2021,Nie2023,2024vanEssen,geest23a,deKeijzer24a}, such as the detection of electron dynamics \cite{Heide2022coherence,Hohenleutner2015}, coherent phonons \cite{Zhang2024}, and material phase transitions \cite{Bionta2021,Nie2023} with in principle sub-cycle (i.e. few to sub-fs) temporal resolution \cite{Hohenleutner2015,garg16a}. Although recent studies have mostly focused on homogeneous samples there is an exciting prospect in spatially resolving these dynamics in systems such as semiconductor devices, nanophotonic metasurfaces, and microscopic flakes. Complicating the imaging of these devices is the fact that their feature size can be well below the wavelength of visible light. The diffraction-limited nature of light in combination with the relatively long wavelengths required for HHG in solids means that simply combining HHG with conventional microscopy will not yield the required resolution for imaging these microscopic systems.\\
To overcome the diffraction limit, super-resolution imaging techniques have been developed, which are widely used in bio(medical)-imaging \cite{2023Sun}. Examples of super-resolution techniques include stimulated depletion microscopy (STED) \cite{Hell1994,2017Blom}, photoactivated localization microscopy (PALM) \cite{Hess2006}, and stochastic optical reconstruction microscopy (STORM) \cite{Xu2017_STORM}. These techniques rely on specific properties of fluorescence, as such they can not simply be applied to HHG imaging. \\
Recently the first super-resolution imaging technique based on the properties of HHG from solids was demonstrated \cite{Murzyn2024}. This harmonic depletion microscopy (HADES) shares similarities with STED as HHG emission is inhibited by an orbital-angular-momentum carrying pre-excitation pulse.
Here we will exploit the strong dependence of the HHG process on the specific properties of the driving field to spatially confine HHG below the diffraction limit in a material-independent and thus generally applicable way.\\
We propose to obtain a high spatial resolution for HHG imaging using ellipticity to control emission via spatial polarization gating (SPG). SPG is inspired by attosecond (temporal) polarization gating \cite{Corkum1994}, the first technique that was proposed to generate isolated attosecond pulses in the extreme-ultraviolet spectral range via HHG from gasses. In polarization gating, a half-cycle overlap of two counter-rotating circularly polarized femtosecond pulses facilitates temporally confining HHG and thus enables the generation of isolated attosecond pulses \cite{Sansone2006}. Spatial confinement of third-harmonic generation using a driver with spatially varying ellipticity has been demonstrated \cite{Masihzadeh2009}.  Here we show general applicability of SPG to HHG in solids, provide a simple framework that allows us to predict the spatial emission profile, and propose an extension of SPG to widefield imaging.

We measured the ellipticity response of HHG from ZnO, silicon, and sapphire (for details see \cite{SM}). In all cases, HHG suppression can be described by a Gaussian, matching with the theoretical results of Ref. \cite{Liu2016} and qualitatively consistent with HHG in gasses \cite{Moller2012},
\begin{equation}
    \frac{I(\epsilon)}{I(0)} = S(\epsilon) = e^{-\frac{1}{2}\left(\frac{\epsilon}{\epsilon_0}\right)^2}.
\end{equation}
Here $\epsilon$ is the ellipticity (for detailed definition see \cite{SM}), and $\epsilon_0$ specifies the efficiency of suppression. $\epsilon_0$ is larger for higher orders ($\epsilon_0 \mathrm{(H3)} = 0.32\pm0.02$; $\epsilon_0 \mathrm{(H5)} = 0.23 \pm 0.02$) but shows little to no intensity and material dependence. This shows that the field characteristics are the dominant factor in the suppression. A simple analytical model to predict the harmonic emission considers the harmonic intensity projected along the polarization direction $\hat{\textbf{r}}$,
\begin{equation}
    I_{n,\hat{\textbf{r}}}(\epsilon)   = |\mathbf{E}_n\cdot\hat{\textbf{r}}|^2
    \propto    S(\epsilon)    \frac{|\mathbf{E}_0\cdot\hat{\textbf{r}}|^{2}}{|\mathbf{E}_0|^{2}}|\mathbf{E}_0|^{2n} .
    \label{eq:analyticalModel}
\end{equation}
Here $n$ denotes the harmonic order, and $\mathbf{E}_0$ is the incoming electric field. The first factor is the suppression factor $S(\epsilon)$. The second term results from the assumption that the polarization of the HHG emission is the same as that of the incoming light, this assumes perfect momentum conservation between the incoming and emitted light and is typically realized for amorphous and polycrystalline samples. The third term describes the intensity scaling of the harmonic with the incoming intensity, which we treat by perturbation theory. This assumption will fail for higher intensities and orders where non-perturbative (and typically lower) intensity scalings are observed \cite{Goulielmakis2022}. For the experiments in this letter, these assumptions were found to hold up. \\
To support the generality of the observations discussed, we have performed simulations of the ZnO ellipticity response by solving the semiconductor Bloch equations (SBE) \cite{Silva2019, wannier-chapter2024}. We have used a simple tight-binding model within the simulations to describe the ZnO (for details see \cite{SM}). Here, the ellipticity is varied by delaying two equally intense 2000 nm pulses with linear and orthogonal polarizations. For these scans, we isolate emission with a polarizer that is aligned diagonally with the s and p directions. \\
The simulation results are shown in Fig. \ref{fig:SBE_Delay}a where we observe periodic suppression of all the harmonics orders.  As the delay is scanned over the length of one fundamental wavelength, the driver ellipticity switches from diagonal linear (maximum HHG) to right-handed circular (full HHG suppression) to anti-diagonal linear (maximum HHG, linear polarization of driver rotated by 90$^\circ$ with respect to diagonal) to left-handed circular (full HHG suppression) and back to diagonal linear. As we filter out the anti-diagonal emission this leaves us with emission maxima when the delay is a multiple of the wavelength, which is what we observe in our simulation results. 
For longer delays partial pulse overlap results in both reduced HHG emission for linear and reduced suppression for circular polarization. 
We compare our SBE simulations with measurements of ZnO, where the ellipticity has similarly been varied by delaying two orthogonally polarized pulses. We observe the same behavior as in the SBE simulation in our measurements in Fig. \ref{fig:SBE_Delay}b: strong enhancement and suppression with a periodicity matching that of the driving wavelength. The pulse duration in the experiments is longer than the one used in the SBE simulation, resulting in less intensity change over the delay scan and spectrally narrower HHG emission. The lines observed in the H7 data can be attributed to the broadband bandgap fluorescence found in ZnO \cite{Rodnyi2011}, the underlying multiphoton excitation of which also exhibits a strong ellipticity dependence. Similar to the SBE simulation, we note a spectral tilt in the harmonic emission maxima, which is due to the difference in phase between the different spectral components in the driving pulse for a given delay. Since this depends on the phases of the driver, the spectral tilt is found consistently between all the harmonics, as illustrated via the dashed white lines which all have the same slope.\\
To better compare the SBE simulation and measurement, their integrated harmonic intensity is shown in Fig. \ref{fig:SBE_Delay}c, together with the analytical model (for H3-5) using equation \ref{eq:analyticalModel}. We observe good agreement between SBE simulation, measurement, and analytical model. The different periodicity between SBE simulation and measurement are caused by a slightly longer than 2000 nm experimental wavelength; this can be deduced from the central wavelength of the harmonics shown in figure \ref{fig:SBE_Delay}b. The fact that our simulations can reproduce our experimental results while using a tight-binding model instead of specific material properties indicates that even in solids the decreased HHG efficiency for increased ellipticity is a general effect that is strongly governed by the properties of the driving field.\\

\begin{figure}
    \centering
    \includegraphics[width=\linewidth]{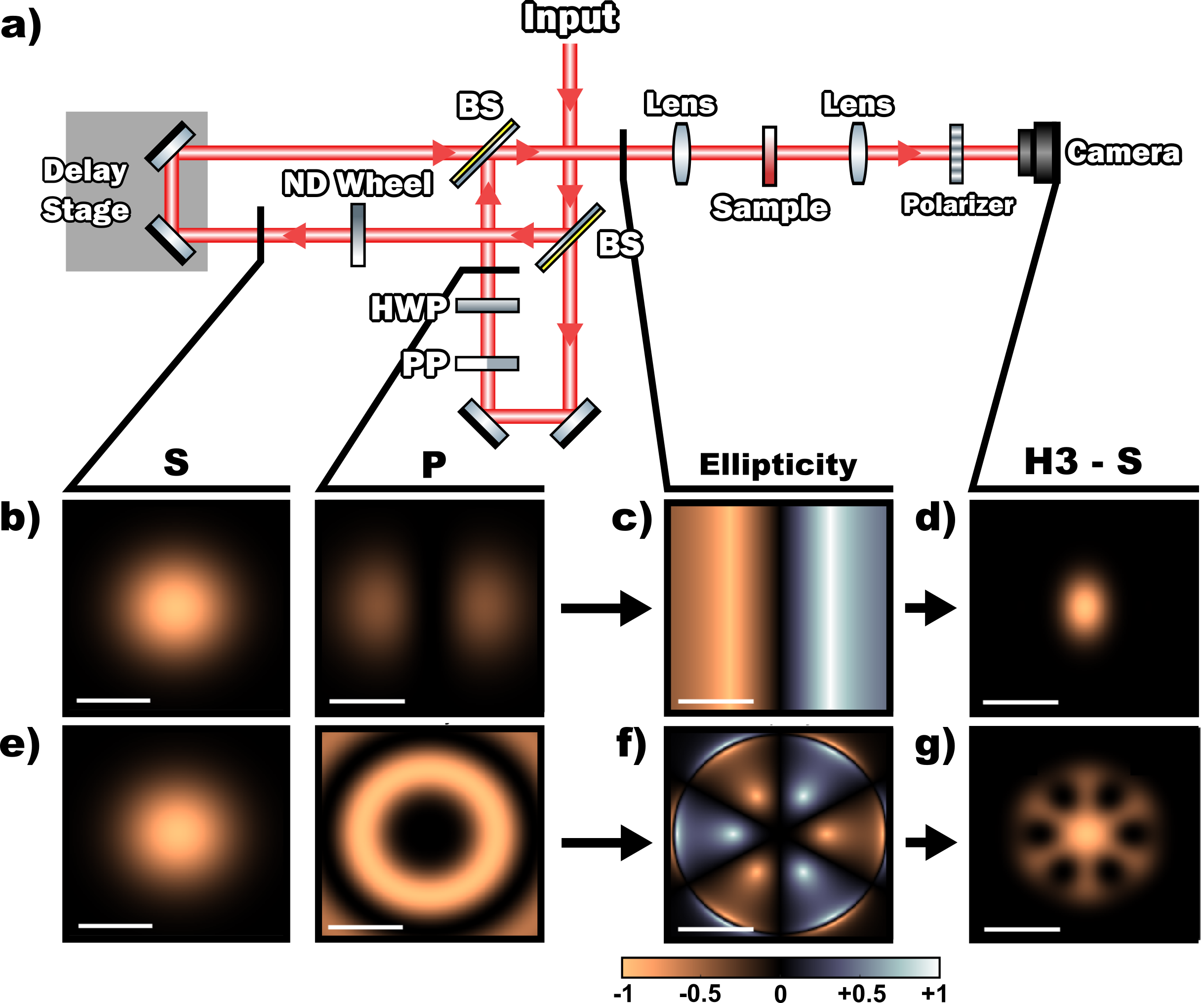}
    \caption{
    (a) Setup for SPG. The input beam is separated by a beamsplitter (BS). In one of the arms, the spatial profile of the beam is changed via a phase plate (PP) and its polarization is changed with a half-wave plate (HWP). Subsequently, the beams are recombined and focused onto the sample. A delay stage and attenuation wheel are used to tune the relative phase and intensity between the two arms. The harmonic emission is imaged with a camera. The polarization of the detected emission is filtered by a polarizer. 
    (b-g) Calculated examples of SPG. 
    (b,e) show profiles of the beams in the two arms. The spatial profiles of the p-polarized fields are (b) HG$_{10}$ and (e) LG$_{31}$.
    (c,f) show the ellipticity after the recombination of the two beams. 
    (d,g) show the H3 emission projected onto the s-polarization direction.
    }
    \label{fig:SPG_setup}
\end{figure}
\begin{figure}
    \centering
    \includegraphics[width=\linewidth]{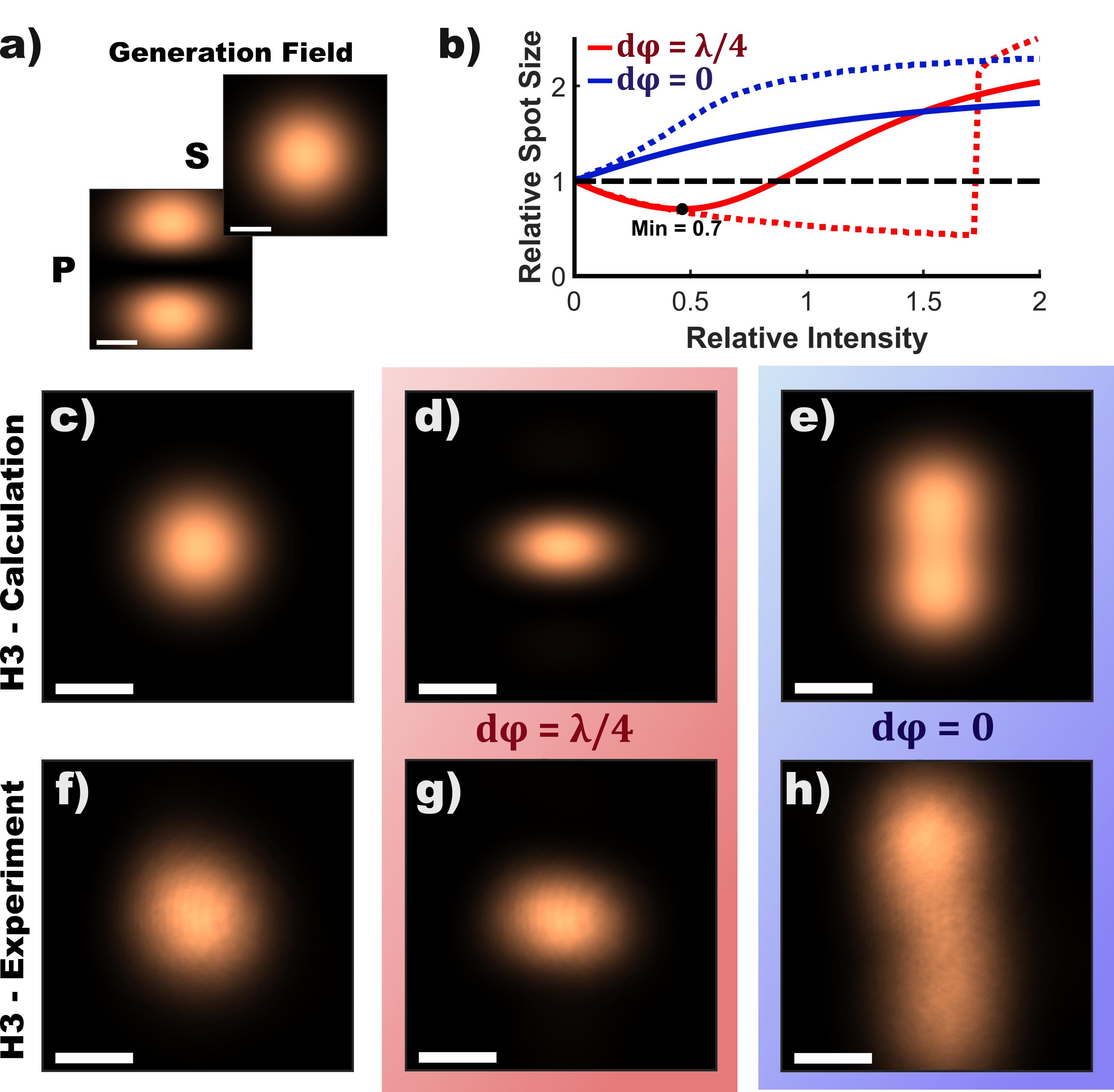}
    \caption{Calculations (a-e) and measurements (f-h) of SPG using an 1800 nm driver where the s-polarization component has a Gaussian profile while the p-polarization component has an HG$_{01}$ profile. 
    (a) Spatial intensity profile of the s and p components.
    (b) H3 spot size for increasing intensity of the p generation field relative to the s generation field, calculated via the image moment (solid) and FWHM (dashed).
    (c-e) Calculated H3 emission spots. (f-h) Corresponding measured H3 emission spots from ZnO.
    White scale bars indicate 30 $\mu$m.
    (c,f) show emission when only the s field is present.
    (d,g) show emission when the s and p fields are $\pi/2$ out of phase.
    (e,h) show emission when the s and p fields are in phase.
    }
    \label{fig:SPG_results}
\end{figure}

So far we have considered HHG by drivers with uniform ellipticity, now we will move to SPG, where we consider beams with spatially varying ellipticity.
We make use of the setup as shown in Fig. \ref{fig:SPG_setup}a. An incoming beam is separated into two arms in one of which the polarization is rotated using a half-wave plate (HWP). The relative spatial profile of the beams in the two arms can now be altered with the use of phase plates, which, when recombining the beams, enables the creation of beams with spatially varying ellipticity. In our calculations and experiments, we have focused on propagation stable excitation profiles, in particular, Hermite-Gaussian (HG$_{mn}$, Fig. \ref{fig:SPG_setup}b-d) and Laguerre-Gaussian (LG$_{lp}$, Fig. \ref{fig:SPG_setup}e-g) modes \cite{Kimel1993}. HG profiles have rectangular symmetry and a phase profile consisting of discrete steps, while the LG profiles have radial symmetry with a phase profile that continuously varies in the azimuthal direction.\\
To evaluate the spatial HHG emission profiles, we use the analytical model described by equation \ref{eq:analyticalModel} and use the Gaussian suppression with the experimentally found parameters for H3 and H5. Examples of calculated SPG emission profiles are shown in Fig. \ref{fig:SPG_setup}d,g. While Fig. \ref{fig:SPG_setup}d shows a narrow HHG emission spot, Fig. \ref{fig:SPG_setup}g illustrates the possibility of creating more complex emission patterns. These calculations clearly show HHG emission being minimal at the positions of maximal ellipticity.\\
To illustrate the capabilities of SPG for imaging we demonstrate spot size reduction below that of conventional HHG emission from ZnO using an 1800-nm driver. For this, we will focus on the H3 emission from the combination of an s-polarized fundamental Gaussian profile with a p-polarized HG$_{01}$ profile, as shown in Fig. \ref{fig:SPG_results}a. The HG$_{01}$ profile is obtained by using a phase step plate with the $\pi$ phase jump centered on the optical axis. The alignment of two beams was optimized to ensure overlap and parallel propagation after recombination. Similar to the delay scans discussed before, the phase of the two beams was scanned using a delay stage; however, instead of recording the emission spectrum, the HHG emission profile was recorded using a camera. Additionally, the polarizer before the detector was rotated to be parallel with the s-polarization.\\
We used the analytical model to evaluate the expected H3 emission spot size, which is shown in Fig, \ref{fig:SPG_results}b. We use two different metrics to evaluate the spot size, the full width at half maximum (FWHM) and the image moment (for details see \cite{SM}). The relative intensity is the intensity of the p component of the driver compared to the s component. We look at the two extremes of the two fields being in-phase (blue) and $\lambda/4$ out-of-phase (red). The in-phase spot size will simply increase with the relative intensity as the effective field strength at the edges of the spot is increased. For the out-of-phase case, we see that a reduction of the spot size can be achieved, which is caused by the edges of the spot becoming elliptically polarized. Comparing the image moment (solid line) and FWHM (dashed line), we see that for the image moment, the spot size reduction only happens for a range of relative intensities, while for the FWHM, the spot size decreases for much longer until it shoots up drastically. For the higher relative intensities, the ellipticity at the edge will be reduced, resulting in the formation of side peaks. These side peaks will result in the HHG emission being emitted throughout a bigger spatial region, however, the central peak still becomes narrower. When the intensity of the side peaks first exceeds half the intensity of the main peak the FWHM spot size is instantly greatly increased.\\
Figure \ref{fig:SPG_results}c-e shows the H3 emission spots for a relative intensity of 1 as calculated (c)-(e) and as measured (f)-(h). In figure \ref{fig:SPG_results}c and e, the emission from only the Gaussian s-polarized driver is shown, which corresponds to conventional Gaussian emission profiles. The spot size of the calculation was chosen to be in good agreement with the measurement. Good agreement is found between the calculations and experimental results for both the out-of-phase and in-phase spots. The out-of-phase spots are considerably narrower in the y-direction, showing the increased spatial confinement of HHG emission. The out-phase spots instead show the greatly increased spot sizes. The maximum experimental spot size reduction found, evaluated via the image moment, was 30\%, close to the predicted minimum.
This demonstrates that our analytical model can predict the spatial HHG emission and that SPG enables localization of emission beyond that of conventional solid HHG emission.\\
The observed spot size reduction demonstrated here is a noticeable improvement, however, bigger improvements are desirable. Inherent to SPG is the need to balance the amplitude and phase of the electric field to achieve maximum ellipticity. To achieve sharp jumps in ellipticity, either sharp phase or amplitude jumps should be present, both of which are diffraction-limited. This means that the local confinement of HHG emission with SPG is still diffraction-limited. It is possible to make use of higher-order harmonics to improve the resolution further (see \cite{SM}). However, this does not solve this inherent limitation.\\
\begin{figure}
    \centering
    \includegraphics[width=\linewidth]{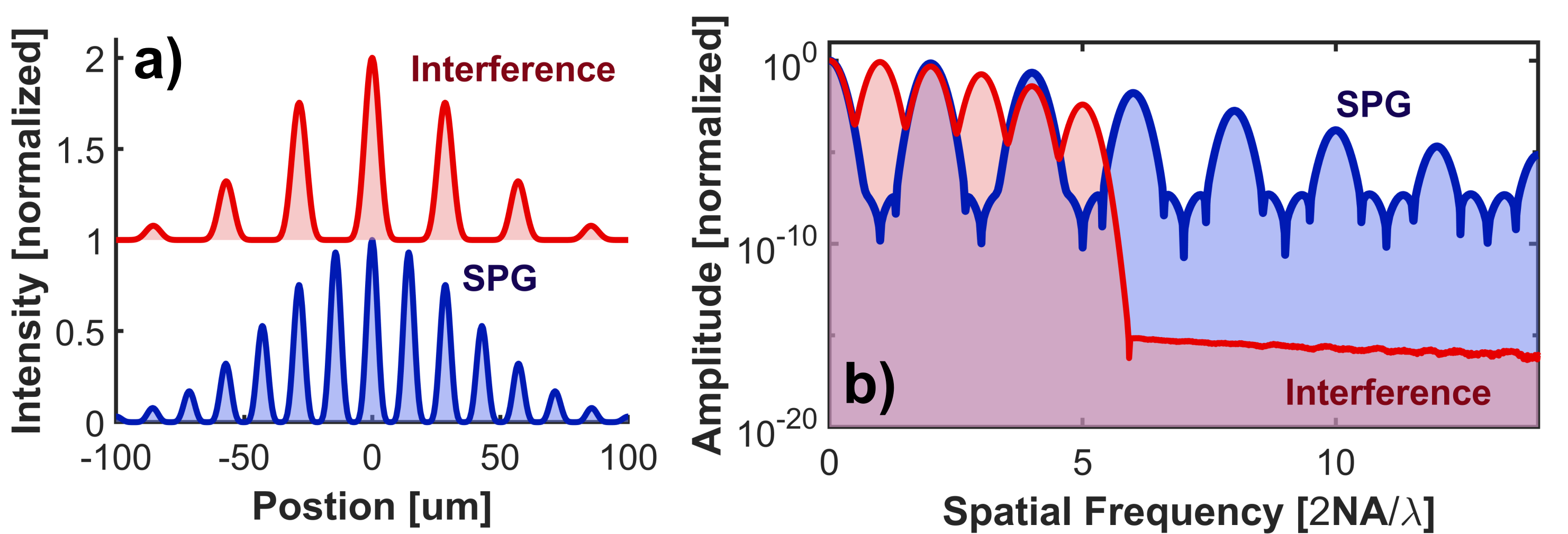}
    \caption{Calculations of structured illumination of H5 with an interference grating and with an ellipticity grating for the same numerical aperture and beam angles. (a) Real space intensity of H5. (b) Spatial frequencies of the emission, $\lambda$ is the driving wavelength.}
    \label{fig:SPGSIM}
\end{figure}

To fully utilize SPG, we will have to move beyond point scanning and instead have to consider widefield techniques, such as structured illumination microscopy (SIM) \cite{Gustafsson2005, Heintzmann2017,Wu2018, Sun2023}. The key here is the observation that although the spatial confinement of HHG emission by SPG is limited, SPG does enable the creation of very sharp, predictable emission features. Conventional SIM uses frequency components in spatially structured illumination patterns to encode information about higher spatial frequencies of the sample into lower spatial frequency components (for details see \cite{SM}). Conventional SIM can increase the imaging resolution to twice the diffraction limit but is essentially still a diffraction-limited imaging method. However, it is possible to enhance the resolution of SIM beyond this by using non-linear processes that introduce higher spatial frequency components into the illumination patterns. 
One way of achieving this is by using the saturating intensity scaling of fluorescence or harmonic generation which is then referred to as saturated structured illumination microscopy (SSIM) \cite{Gustafsson2005,Field2016}. By using SPG to create the illumination patterns in SIM (SPG-SIM), we can instead use the non-linear ellipticity scaling to introduce higher spatial frequencies. SPG-SIM enables super-resolution imaging without needing the high intensities required to reach the saturating regime, alternatively, SPG-SIM can be used in conjunction with SSIM.\\
To illustrate the super-resolution capabilities of SPG-SIM, Fig. \ref{fig:SPGSIM} shows the calculated H5 emission of 2000 nm from respectively, a conventional interference grating pattern and an SPG ellipticity-grating pattern. Both gratings are created by the overlap of two 2000 nm pulses, which have an angle between them (NA$=0.125$). The difference between the two gratings is that for the conventional grating, the beams have the same polarization while for the ellipticity grating they have orthogonal linear polarization. In SPG, the change from linear to circular happens within a half-cycle of the field, instead of a full cycle, this difference is reflected in Fig. \ref{fig:SPGSIM} where the SPG emission profile has twice the number of peaks. The Fourier-transformed emission patterns are shown in Fig. \ref{fig:SPGSIM}b. The interference grating shows a sharp cut-off at the diffraction limit of H5 as expected. The SPG pattern instead has spectral components exceeding this sharp cut-off indicating its capabilities to be used for super-resolution imaging. Realistic resolution improvements are difficult to deduce from these calculations as these will now depend on the noise in the imaging system. If we consider that any amplitude contributions above $10^{-5}$ can be detected then SPG-SIM doubles the effective spatial frequencies in the illumination pattern which increases the maximum spatial frequency that can be detected from twice the diffraction for conventional SIM to three times the diffraction limit (for details see \cite{SM}). For this low NA example, the diffraction limit of H5 is 1600 nm making the resolution limit of conventional SIM 800 nm, while the resolution of SPG-SIM goes up to about 530 nm. This relative increase stays consistent for a different NA, i.e. at an NA of 1, a resolution of 67 nm is possible. These calculations support the possibility of using SPG to enable wide-field super-resolution imaging techniques for solids.\\

In conclusion, we introduced SPG for confining HHG below the diffraction limit, which can find application for high-resolution HHG imaging. Important for SPG is the common ellipticity response in solids where strong suppression of HHG emission is observed, qualitatively matching the atomic response. This ellipticity response was here qualitatively reproduced with SBE simulations without requiring detailed modeling of our material, indicating a dominance of the field characteristics. Spatial confinement of HHG by SPG can be predicted using a simple analytical framework. We demonstrated an H3 spot size reduction of about 30\%, closely matching our calculated optimum. Exciting opportunities lie in applying SPG to higher-order harmonics and combining it with structured illumination. With these next steps, SPG can pave the way for high temporal and spatial resolution imaging in solids.\\

\section{Acknowledgments}
\begin{acknowledgments}
This work has been carried out at the Advanced Research Center for Nanolithography (ARCNL), a public-private partnership of the University of Amsterdam (UvA), the Vrije Universiteit Amsterdam (VU), the Netherlands Organisation for Scientific Research (NWO), and the semiconductor equipment manufacturer ASML, and was partly financed by ‘Toeslag voor Topconsortia voor Kennis en Innovatie (TKI)’ from the Dutch Ministry of Economic Affairs and Climate Policy. 
This manuscript is part of a project that has received funding from the European Research Council (ERC) under the European Union’s Horizon Europe research and innovation programme (grant agreement no. 101041819, ERC Starting Grant ANACONDA) and funded P.J.v.E. and partly P.M.K.
The manuscript is also part of the VIDI research programme HIMALAYA with project number VI.Vidi.223.133 financed by NWO, which partly funded P.M.K.
P.M.K. acknowledges support from the Open Technology Programme (OTP) by NWO, grant no. 18703.
E. B. M. and R. E. F. S. acknowledge support from the fellowship LCF/BQ/PR21/11840008 from \textquotedblleft La Caixa\textquotedblright{} Foundation (ID 100010434). This research was supported by Grant PID2021-122769NB-I00 funded by MCIN/AEI/10.13039/501100011033. A.J.G. acknowledges support from Comunidad de Madrid through TALENTO Grant 2022-T1/IND-24102.

\end{acknowledgments}

\newpage
\onecolumngrid
\section{Supplementary Material}
\subsection{I - Ellipticity Response of High-Harmonic Generation from Solids}
\begin{figure}
    \centering
    \includegraphics[width=0.6\linewidth]{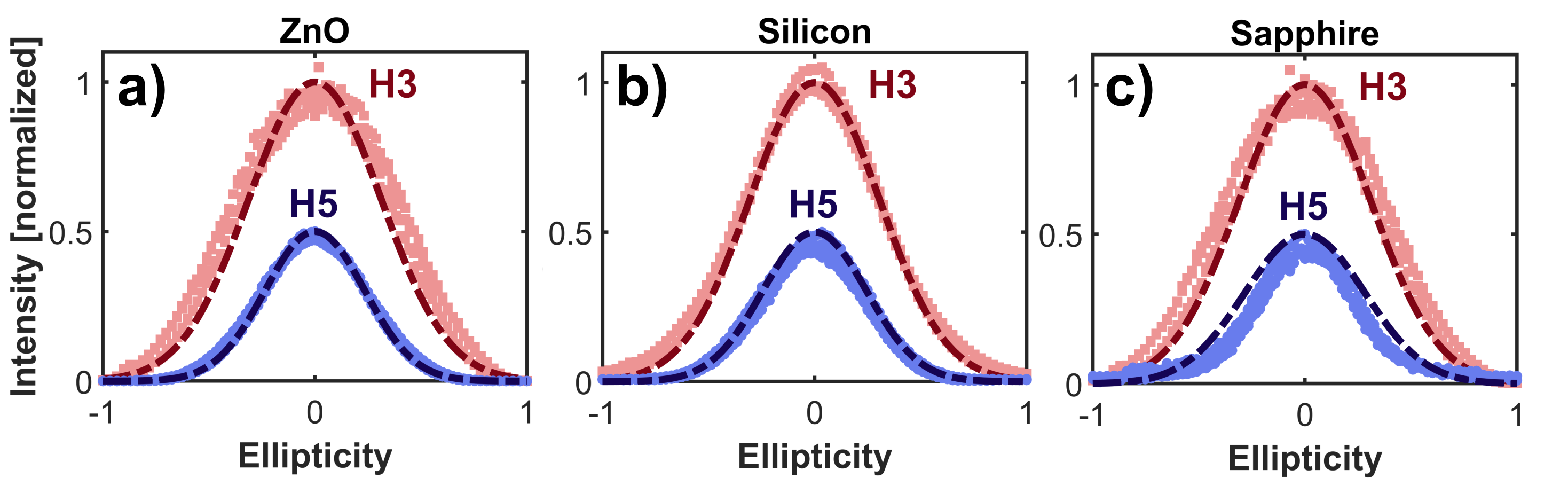}
    \caption{(a)-(c) show the ellipticity response of the H3 (red) and H5 (blue) intensity, respectively, measured from (c) silicon thin film, (d) ZnO, and (e) sapphire. The ellipticity of the driver has been varied using a quarter-wave plate. The H5 intensity has been normalized to 0.5 to improve readability. The dashed lines show the Gaussian curves with the moments ($\epsilon_0$) corresponding to those found via the center of mass calculations of the respective curves.}
    \label{fig:EllipticityResponse}
\end{figure}

The ellipticity response of HHG in gasses has long been investigated \cite{Antoine1996,Moller2012}. For atomic gasses, the HHG emission decreases rapidly when the driving field's ellipticity is increased. More recent studies have investigated the ellipticity response of solid HHG \cite{Liu2016,Zhang2019,Tamaya2016, Hollinger2021, 2017TancogneDejean, Heide2022, Klemke2020, Sekiguchi2023,You2017, Yoshikawa2017}. Solids introduce additional complexity which has translated to the observation of anisotropic ellipticity responses and increased HHG emission for elliptical excitation both of which were found to strongly depend on the topology of the material, the intensity of the excitation field, and the observed harmonic order \cite{Heide2022, Sekiguchi2023, Yoshikawa2017}. In this work, we will instead focus on the big class of solids with weak anisotropy where the ellipticity response resembles the atomic response.\\
To illustrate this common ellipticity response we measured the HHG emission generated by an 1800 nm driver from a number of semiconductor materials: ZnO, Silicon (thin film), and Sapphire. Figure \ref{fig:EllipticityResponse} shows the measured ellipticity response of harmonics H3 (600 nm) and H5 (360 nm), where the ellipticity of the driver was varied using a quarter-wave plate (QWP). For all the different materials clear suppression of the HHG emission is observed with increased ellipticity. 

\subsection{II - Ellipticity}
\begin{figure}
    \centering
    \includegraphics[width=0.6\linewidth]{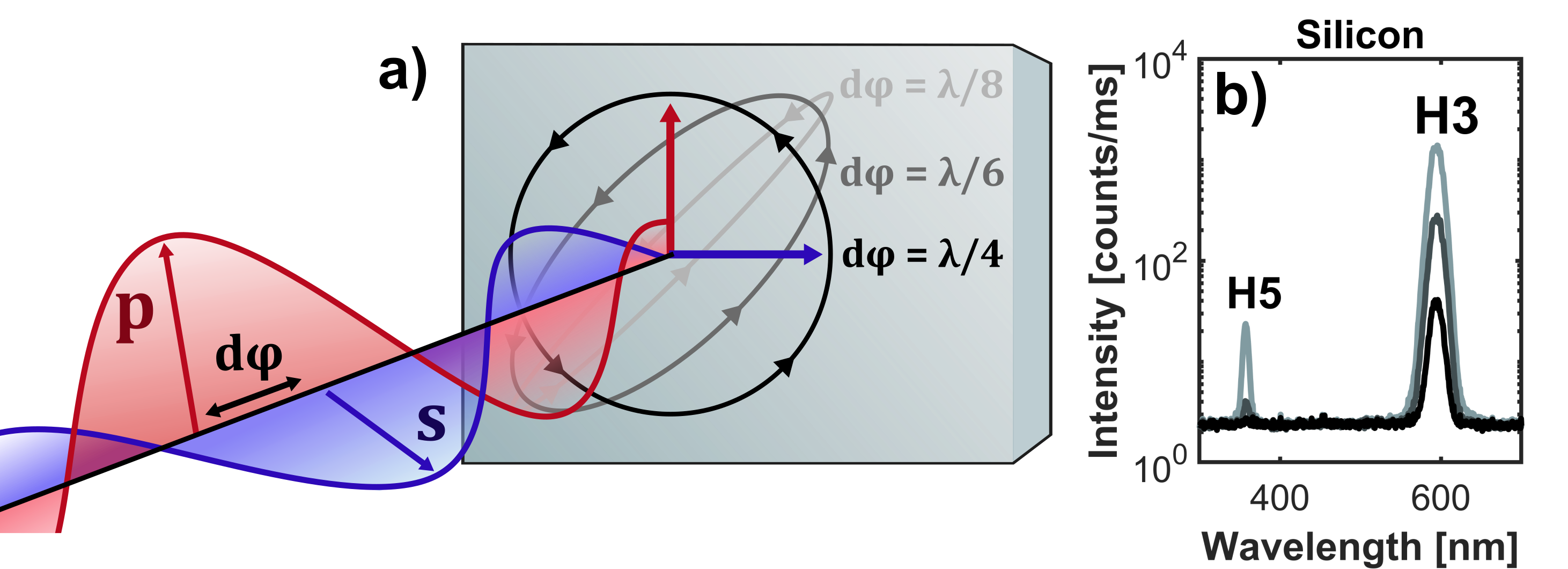}
    \caption{(a) shows how elliptically polarized light can be described as the combination of two orthogonal linearly polarized fields with a phase $\Delta\phi$. (b) shows the HHG spectra from silicon thin film for illumination with increasing ellipticity where the ellipticity was varied using a quarter-wave plate.}
    \label{fig:EllipticityDecomposition}
\end{figure}

The ellipticity is defined as the ratio of the field amplitudes along the major and minor axis, while the sign of the ellipticity is dependent on the relative phase of these components:
\begin{equation}
    \epsilon = \abs{\frac{E_{\text{minor}}}{E_{\text{major}}}} \cdot
    \begin{cases}
     \frac{\text{sin}(\Delta \phi)}{\abs{\text{sin}(\Delta \phi)}} &\text{ if } \quad\Delta \phi \neq 2\pi \text{n}.\\
    0 &\text{ otherwise }
    \end{cases}.
\end{equation}
The major and minor axis are, respectively, the directions in which the field amplitude is maximal and minimal. The major and minor axis are orthogonal, and the fields in these directions have $\lambda/4$ phase differences. The ellipticity $\epsilon$ is 0 for linear polarized light and $\pm1$ for circularly polarized light. The sign of $\epsilon$ indicates left versus right-handed polarization. As a QWP is used to vary ellipticity, the orientation of the wave plate determines the relative field strengths projected onto the major and minor axes. In our experiments, we also create elliptically polarized light by combining two orthogonal fields, for this we have to consider the decomposition of the total electric field in these two orthogonal components $\mathbf{E}_s$ and $\mathbf{E}_p$, as also shown in Fig. \ref{fig:EllipticityDecomposition}. In this framework, the ellipticity can be given as a function of the relative amplitude and phase.\\
To evaluate the ellipticity we consider the orthogonal field components $\mathbf{E}_s = E_s\text{sin}(\omega t)$ and $\mathbf{E}_p = E_p\text{sin}(\omega t + \Delta\phi)$, these define the total field:
\begin{equation}
    \mathbf{E} = E_s \text{sin}(\omega t) \hat{x} + E_p\text{sin}(\omega t + \Delta\phi)\hat{y}.
\end{equation}
To obtain the major and minor axes, we have to consider the amplitude of this field,
\begin{equation}
    \abs{\mathbf{E}}^2 = E_s^2\text{sin}^2(\omega t)+ E_p^2\text{sin}^2(\omega t + \Delta\phi).
\end{equation}
We are interested only in the ratio of the major and minor axes so we can reduce our number of parameters by considering the ratio $A = E_s/E_p$,
\begin{equation}
    \frac{\abs{\mathbf{E}}^2}{E_p^2} = A^2\text{sin}^2(\omega t)+ \text{sin}^2(\omega t + \Delta\phi).
    \label{eq:axis}
\end{equation}
To obtain the major and minor amplitudes the extremes of this equation have to be found. The local extremes are found at,

\begin{equation}
\begin{split}
        t = & \text{ tan}^{-1}\biggl(\frac{1}{2} \text{cot}(\Delta\phi) \Bigl(A^2 \text{tan}^2(\Delta\phi) + A^2\\ 
        &- \sqrt{(\text{tan}^2(\Delta\phi) + 1) (A^4 \text{tan}^2(\Delta\phi) + A^4 - 2 A^2 \text{tan}^2(\Delta\phi) + 2 A^2 + \text{tan}^2(\Delta\phi) + 1)}\\
        & - \text{tan}^2(\Delta\phi) + 1 \Bigl) \biggl) + \frac{\pi}{2}\text{n}_1, \text{ if } \Delta\phi \neq \frac{\pi}{2}\text{n}_2.
\end{split}
\end{equation}
We can plug this expression back into equation \ref{eq:axis} to obtain,
\begin{equation}
    \begin{split}
        \frac{\abs{\mathbf{E}}^2}{E_p^2}\biggl|_\text{minor} = & \text{ }\Biggl(A^2 \text{cot}^2(\Delta\phi) \biggl(A^2 \text{tan}^2(\Delta\phi)+A^2 -\text{tan}^2(\Delta\phi)+1\\
        &-\sqrt{(\text{tan}^2(\Delta\phi)+1) (A^4 \text{tan}^2(\Delta\phi)+A^4-2 A^2 \text{tan}^2(\Delta\phi)+2 A^2+\text{tan}^2(\Delta\phi)+1)} \biggl)^2 \Biggl)\\
        &\cdot \Biggl(4 \biggl(\frac{1}{4} \text{cot}^2(\Delta\phi) \biggl(A^2 \text{tan}^2(\Delta\phi)+A^2 -\text{tan}^2(\Delta\phi)+1 \\
        &- \sqrt{(\text{tan}^2(\Delta\phi)+1) (A^4 \text{tan}^2(\Delta\phi)+A^4-2 A^2 \text{tan}^2(\Delta\phi)+2 A^2+\text{tan}^2(\Delta\phi)+1)}\biggl)^2+1\biggl) \Biggl)^{-1}\\
        &+ \text{sin}^2\Biggl(A \text{tan}\biggl(\frac{1}{2}\text{cot}(\Delta\phi) \biggl(A^2 \text{tan}^2(\Delta\phi)+A^2 -\text{tan}^2(\Delta\phi)+1\\
        &- \sqrt{(\text{tan}^2(\Delta\phi)+1) (A^4 \text{tan}^2(\Delta\phi)+A^4-2 A^2 \text{tan}^2(\Delta\phi)+2 A^2+\text{tan}^2(\Delta\phi)+1)}
        \biggl)\biggl)+\Delta\phi \Biggl),\\
    \end{split}
\end{equation}
\begin{equation}
    \begin{split}        
         \frac{\abs{\mathbf{E}}^2}{E_p^2}\biggl|_\text{major} = &  A^2 \Biggl(\frac{1}{4} \text{cot}^2(\Delta\phi) \biggl(A^2 \text{tan}^2(\Delta\phi)+A^2-\text{tan}^2(\Delta\phi)+1\\
         &-\sqrt{(\text{tan}^2(\Delta\phi)+1) (A^4 \text{tan}^2(\Delta\phi)+A^4-2 A^2 \text{tan}^2(\Delta\phi)+2 A^2+\text{tan}^2(\Delta\phi)+1)}\biggl)^2+1\Biggl)^{-1}\\
         &+   \text{cos}\Biggl(A\text{tan}\biggl(\frac{1}{2} \text{cot}(\Delta\phi) \biggl(A^2 \text{tan}^2(\Delta\phi)+A^2-\text{tan}^2(\Delta\phi)+1\\
         &-\sqrt{(\text{tan}^2(\Delta\phi)+1) (A^4 \text{tan}^2(\Delta\phi)+A^4-2 A^2 \text{tan}^2(\Delta\phi)+2 A^2+\text{tan}^2(\Delta\phi)+1)}\biggl) \biggl)+\Delta\phi \Biggl)^2.
    \end{split}
\end{equation}
From this, we can exactly evaluate the ellipticity,
\begin{equation}
    \epsilon = \frac{\text{sin}(\Delta \phi)}{\abs{\text{sin}(\Delta \phi)}} \sqrt{\frac{\abs{\mathbf{E}}^2}{E_p^2}\biggl|_\text{minor}  \left(\frac{\abs{\mathbf{E}}^2}{E_p^2}\biggl|_\text{major}\right)^{-1}} \quad \text{ if } \quad \Delta\phi \neq \frac{\pi}{2}\text{n}.
\end{equation}
Additionally, for two particular cases, we obtain simplified exact expressions. When the field components our out-of-phase ($\Delta\phi = \pi/2 + \pi\text{n}$) the ellipticity is given by:
\begin{equation}
        \epsilon = \frac{\text{sin}(\Delta \phi)}{\abs{\text{sin}(\Delta \phi)}} \text{min}\left(\abs{\frac{E_p}{E_s}},\abs{\frac{E_s}{E_p}} \right).
        \label{eq:outOfPhase}
\end{equation}
In this case, the major and minor axis have the same orientation as the s and p components. When the amplitudes of both fields are equal ($A = 1$) the ellipticity is given by:
\begin{equation}
        \epsilon = \frac{\text{sin}(\Delta \phi)}{\abs{\text{sin}(\Delta \phi)}} \text{min}\left(
        \sqrt{\frac{1+\text{cos}(\Delta\phi)}{1-\text{cos}(\Delta\phi}},
        \sqrt{\frac{1-\text{cos}(\Delta\phi)}{1+\text{cos}(\Delta\phi}} \right).
        \label{eq:sameAmplitude}
\end{equation}

\subsection{III - Experimental Method}
\begin{figure}
    \centering
    \includegraphics[width=0.6\linewidth]{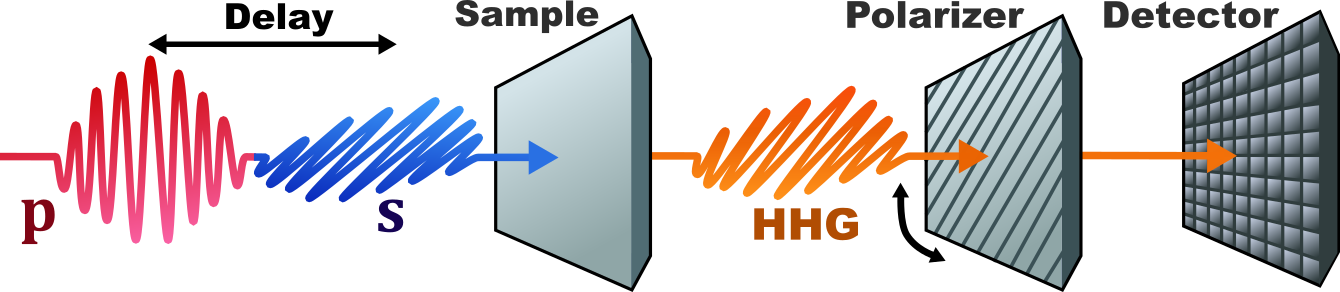}
    \caption{The delay between two orthogonal polarized excitation pulses is varied to change the ellipticity. To obtain near uniform elliptical polarized light the two pulses need to almost completely overlap. For partial overlap of the pulses, only small parts of the pulses can be elliptically polarized. A polarizer before the detector allows the filtering of particular polarization components of the HHG emission. }
    \label{fig:delaySetup}
\end{figure}

In this work we presented results from three distinct measurements, basic ellipticity measurements as shown in Fig. \ref{fig:EllipticityResponse}, delay measurements as shown in main text Fig. 1b, and SPG measurements as shown in main text Fig. 3. In all these measurements we use IR drivers generated by an OPA which we weakly focus to generate HHG with peak intensities between 300 GW/cm$^2$ to 3 TW/cm$^2$. The beam out of the OPA was used as is, without any spatial filtering of the mode profile. All measurements were performed in transmission. \\
The basic ellipticity measurements were performed with an 1800 nm driver of which the ellipticity was varied using a quarter-wave plate. The ellipticity response was measured from ZnO, sapphire, SiN, silicon (thin film), and UV-fused silica.\\
The delay measurements were done with ZnO using a 2000 nm driver of which the ellipticity was varied by delaying two orthogonally polarized beams, as illustrated in Fig. \ref{fig:delaySetup}. The emission was filtered using a polarizer, with the polarizer aligned as to filter out the anti-diagonal emission with respect to the s- and p-polarized driver beams. The relative phase and amplitude between the two beams were respectively varied using a delay stage and a neutral density filter wheel.\\
The SPG measurements were performed with ZnO using a 1800 nm driver with the ellipticity being varied similarly to the delay measurements. However, the polarizer was now aligned to filter out the p-polarized components. Additionally, a phase plate was used in this measurement to alter the spatial profile of the p-polarized beam. 

\subsection{IV -  SBE Simulation}
To simulate the nonlinear response, we numerically solve the Semiconductor Bloch Equations in the Wannier gauge \cite{Silva2019, wannier-chapter2024}. 
We have developed a tight-binding model that captures the relevant properties of ZnO, most importantly the amplitude of the bandgap. The model consists of a two-dimensional square lattice with two atoms, A and B, per unit cell, which is a simplified geometry compared to the hexagonal lattice structure of the ZnO sample. The band gap of the system is equal to the difference of their onsite energies, i.e. $\Delta = |\epsilon_A-\epsilon_B|$. We matched the theoretical gap to the experimental one $\Delta = \Delta_{\text{ZnO}}\approx 3.3$ eV. The model has inversion symmetry, which can also be seen from the absence of even harmonics in the simulations. We only considered hopping between nearest neighbors of the lattice with a magnitude of $t=1.0$ eV $\ll \Delta_\textrm{ZnO}$. The system is driven using two orthogonal pulses with a cos$^2$ envelope duration of 32.5 fs, peak intensity of 4 TW/cm$^2$, and a wavelength of 2 $\mu$m. Consequently, the delay between pulses is scanned using a 0.4 fs step size. Convergence of the simulations is verified to be achieved at 512x512 points of a Monkhorst-Pack grid in reciprocal space and a propagation time step size of 0.097 fs.

\subsection{V - Spot Size}
To evaluate the spot sizes of the HHG emission, we approximate their shape with ellipsoids and consider their enclosed area:
\begin{equation}
    A = \pi R_{\text{max}} R_{\text{min}}.
\end{equation}
Here, $R_{\text{max}}$ and $R_{\text{min}}$ are the long and short radii of the ellipsoid. We evaluate these radii in two different ways either via the image moment or by calculating the full-width half maximum (FWHM) of the x and y cross-sections. To obtain the relative spot size, the evaluated area is normalized to the area of the HHG spot size, which is generated from only the fundamental Gaussian. For Gaussian spots, the image moment and FWHM will yield the same relative spot sizes, For more complexly shaped spots these two metrics will differ. In particular, when having emission profiles with noticeable side peaks, the image moment will yield a relatively big spot size, while the FWHM will disregard these side peaks and yield smaller spot sizes. The image moment provides a metric on the spatial distribution of the intensity profile throughout space, while the FWHM tells us more about the sharpness and size of the high peak intensity central feature.\\
Generally, for images, the momenta can be evaluated using:
\begin{equation}
    M_{n,m} = \sum_x\sum_y x^n y^m I(x,y).
\end{equation}
The mean values can be calculated using:
\begin{equation}
    \bar{\mu}_x = \frac{M_{10}}{M_{00}} \quad \text{and} \quad \bar{\mu}_y = \frac{M_{01}}{M_{00}}
\end{equation}
The second-order normalized mean-corrected momenta are given by,
\begin{equation}
    \begin{split}
        \mu_{20} &= \frac{M_{20}}{M_{00}} - \bar{\mu}_x^2,\\
        \mu_{02} &= \frac{M_{02}}{M_{00}} - \bar{\mu}_y^2,\\
        \mu_{11} &= \frac{M_{11}}{M_{00}} - \bar{\mu}_x\bar{\mu}_y.
    \end{split}
\end{equation}
Note that for Gaussian distributions the x and y standard deviations are respectively given by $\sigma_x = \sqrt{\mu_{20}}$ and $\sigma_y = \sqrt{\mu_{02}}$. From the second-order moment, the major and minor axes are given by,
\begin{equation}
    \lambda_{\pm} = \sqrt{\frac{\mu_{20}+\mu_{02}}{2}\pm\sqrt{4\mu_{11}^2+\frac{(\mu_{20}-\mu_{02})^2}{2}}}.
\end{equation}
These major and minor radii are used as the long and short radii of the ellipsoidal spots. 

\subsection{VI - Spot Size Reduction H5}
\begin{figure}
    \centering
    \includegraphics[width=0.6\linewidth]{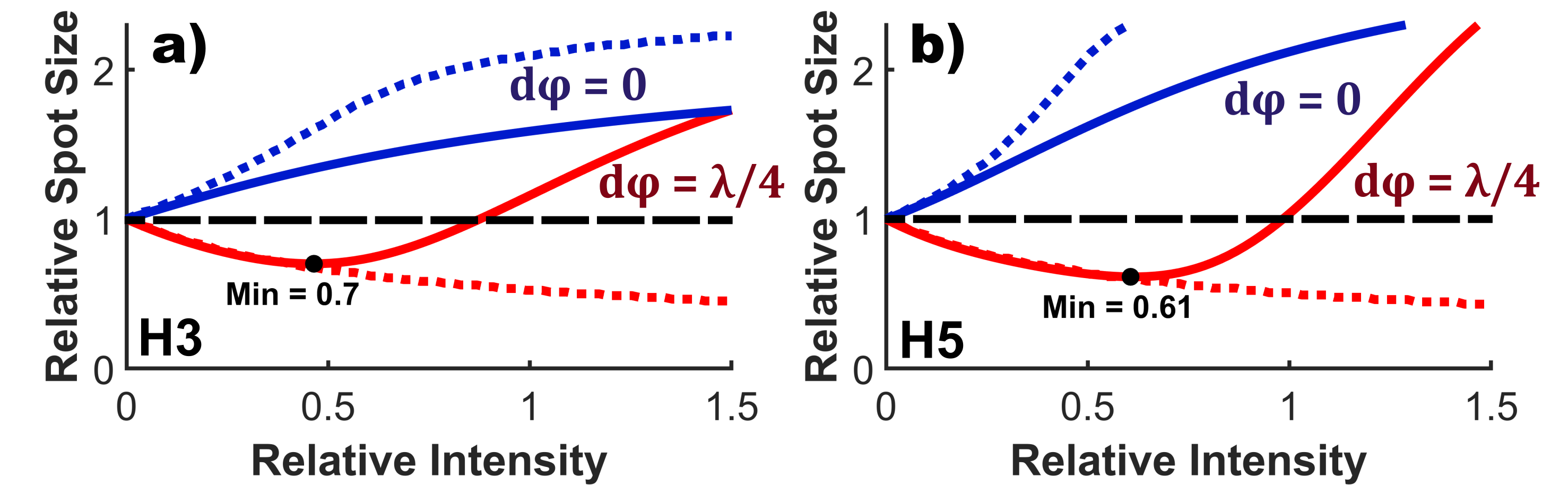}
    \caption{Calculations of SPG where the s-polarization component has a Gaussian profile while the p-polarization component has an HG$_{01}$ profile. The spot size is shown for increasing the intensity of the p-generation field relative to the s-generation field. The spot size is calculated via the image moment (solid) and FWHM (dashed). In (a) for H3 with $\epsilon_0 = 0.32$ (this is the same data as Fig. 3b in the main text, focusing here on the lower relative intensities) and (b) H5 with $\epsilon_0 = 0.23$.}
    \label{fig:Bonus_ResH5}
\end{figure}
The relative spot size reduction that can be achieved is different for the different harmonic orders. To illustrate this the calculated spot size reduction of H3 and H5 are shown in Fig. \ref{fig:Bonus_ResH5}. We see that the optimal spot size reduction evaluated with the image moment increases from about 30\% for H3 to close to 40\% for H5.

\subsection{VII - Structured Illumination Microscopy}
In widefield microscopy, the image detected by a camera $D$ can be described as \cite{Sun2023}:
\begin{equation}
    D(\mathbf{r}) = \Big(s(\mathbf{r})\cdot I(\mathbf{r})\Big)* \text{PSF}(\mathbf{r}).
\end{equation}
Here $s$ is the structure that is being imaged, $I$ is the illumination profile, and PSF is the point-spread function. We here use $*$ to denote a convolution. To evaluate the possible resolution we can consider the detected image in the spatial-frequency domain:
\begin{equation}
    \mathcal{F}\{D\} = \Big(\mathcal{F}\{s\}* \mathcal{F}\{I\}\Big) \cdot \mathcal{F}\{\text{PSF}\}.    \label{eq:SIM_Fourier}
\end{equation}
Here $\mathcal{F}$ indicates the Fourier transform. In conventional microscopes, the PSF will be close to an Airy disk which is close to the shape of a Gaussian, which means that equally the Fourier transform of the PSF will resemble a Gaussian. The multiplication of the PSF in Eq. \ref{eq:SIM_Fourier} functions as a filter for the higher spatial frequencies, which results in diffraction-limited imaging. For conventional widefield imaging, the illumination pattern is homogenous, making retrieving information about the higher spatial frequency components from the structure impossible.\\
In structured illumination microscopy (SIM) a structure is introduced to the illumination pattern that can shift information of the higher spatial frequency components of the structure to lower spatial frequencies. Effectively the structure and illumination pattern together will result in a Moiré pattern. The convolution in Eq. \ref{eq:SIM_Fourier} has frequency components at,
\begin{equation}
    \Omega_s \pm \Omega_I.
\end{equation}
With $\Omega_s$ and $\Omega_I$ being respectively a spatial frequency component of the structure and illumination. If the difference in frequencies is small enough they will not be filtered out by the PSF which enables detection of the higher spatial frequency components of the structure. If the PSF filters out all frequencies above the diffraction-limit $\Omega_{\text{diff}}$, then the maximum frequency of the structure about which information can be detected is
\begin{equation}
    \Omega_{\text{diff}} + \text{max}(\Omega_{I}).
\end{equation}
If the maximum frequency in the illumination pattern has the same diffraction limit as the PSF then the maximum detectable frequency is $2 \Omega_{\text{diff}}$ which is the case for conventional SIM. In the example shown in main text Fig. 4, we see that the SPG illumination pattern has spatial frequencies up to at least twice the diffraction limit of H5 which means that the maximum detectable spatial frequency will be $3 \Omega_{\text{diff,H5}}$, which in terms of real space improvement is a resolution increase of 33\% beyond conventional SIM.

\twocolumngrid
\bibliography{sources}

\end{document}